\documentclass[11pt]{article}
\usepackage{moriond,epsfig}
\usepackage{amssymb}

\def\Journal#1#2#3#4{{#1} {\bf #2}, #3 (#4)}

\def\NIMA{{\em Nucl. Instrum. Methods} A}
\def\NPA{{\em Nucl. Phys.} A}
\def\NPB{{\em Nucl. Phys.} B}
\def\PLB{{\em Phys. Lett.} B}
\def\PRL{\em Phys. Rev. Lett.}

\newcommand {\snn} 	{\sqrt{s_{_{\rm NN}}}}

\newcommand {\zt}       {z_{\rm T}}
\newcommand {\pt}       {p_{\rm T}}
\newcommand {\pttrig}	{p_{\rm T}^{\rm trig}}
\newcommand {\ptasso}	{p_{\rm T}^{\rm assoc}}

\newcommand {\dphi}	{\Delta\phi}

\def\be{\begin{equation}}
\def\ee{\end{equation}}
\def\bea{\begin{eqnarray}}
\def\eea{\end{eqnarray}}

\begin{document}
\vspace*{4cm}
\title{HIGH-PT HADRON PRODUCTION AND TRIGGERED PARTICLE CORRELATIONS}

\author{A. MISCHKE {\em for the STAR Collaboration} \footnote{For the full list of STAR
authors, see reference~\cite{starwhite}.}}

\address{NIKHEF, Amsterdam and Institute for Subatomic Physics,\\ Utrecht University, 
Princetonplein 5, 3584 CC Utrecht, The Netherlands.\\ {\tt Email:a.mischke@phys.uu.nl}.}

\maketitle\abstracts{\vspace*{2mm} The STAR experiment at the Relativistic Heavy-Ion 
Collider has performed measurements of high transverse momentum particle production in 
ultra-relativistic heavy-ion collisions. 
High-$\pt$ hadrons are generated from hard parton scatterings early in the collision. 
The outgoing partons probe the surrounded hot and dense matter through interactions.
Recent results on high-$\pt$ inclusive particle production and leading particle 
correlations in p+p, d+Au and Au+Au collisions at $\snn = 200$ GeV are reviewed.}

\section{Introduction}\label{subsec:prod}
The aim of studying ultra-relativistic heavy-ion collisions is to find evidence for the
production of a new state of strongly interacting matter, the Quark-Gluon Plasma, where
quarks and gluons are deconfined. The Relativistic Heavy-Ion Collider (RHIC) at Brookhaven
National Laboratory provides Au+Au collisions at the highest energy presently available
of $\snn = 200$ GeV. In this energy range parton hard scattering in the initial state 
plays a significant role. The scattered partons can be used to probe the medium produced in
the collision. Theoretical models~\cite{TheoIntro} predict partonic energy loss via medium
induced gluon Bremsstrahlung (jet-quenching) and jet broadening due to parton rescattering
in the extremely dense medium. This parton energy loss is sensitive to the medium density.

Parton energy loss in the medium can be studied by measuring the production of high-$\pt$ 
particle yields and azimuthal correlations. A strong suppression (factor of 4--5) of 
high-$\pt$ hadron production relative to a simple binary collision scaling from proton-proton 
collisions has been observed in central Au+Au collisions~\cite{starAuAuSpectra1,starAuAuSpectra2}. 
Additionally, it was found that jet-like correlations opposite to trigger hadrons are strongly 
suppressed~\cite{starAuAuCorr}. In contrast, no suppression effects were observed in d+Au 
collisions~\cite{stardAuSpectraCorr}, which reflects the effects in cold nuclear matter. These 
results are consistent with the jet-quenching mechanism as a final state effect. 
Further RHIC results are reviewed in references~\cite{starwhite,jacobs}.

In this paper a selection of recent results on hard probes in nuclear collisions from 
the STAR experiment is presented. The STAR experiment~\cite{NimSTAR} has extended its 
high-$\pt$ measurements in a recent high statistics run in 2004 with 30M minimum-bias and 
18M central events.

\section{Inclusive identified particle spectra}
\begin{figure}[t]
\begin{minipage}[t]{60mm}
\hspace{-6mm}
\epsfig{file=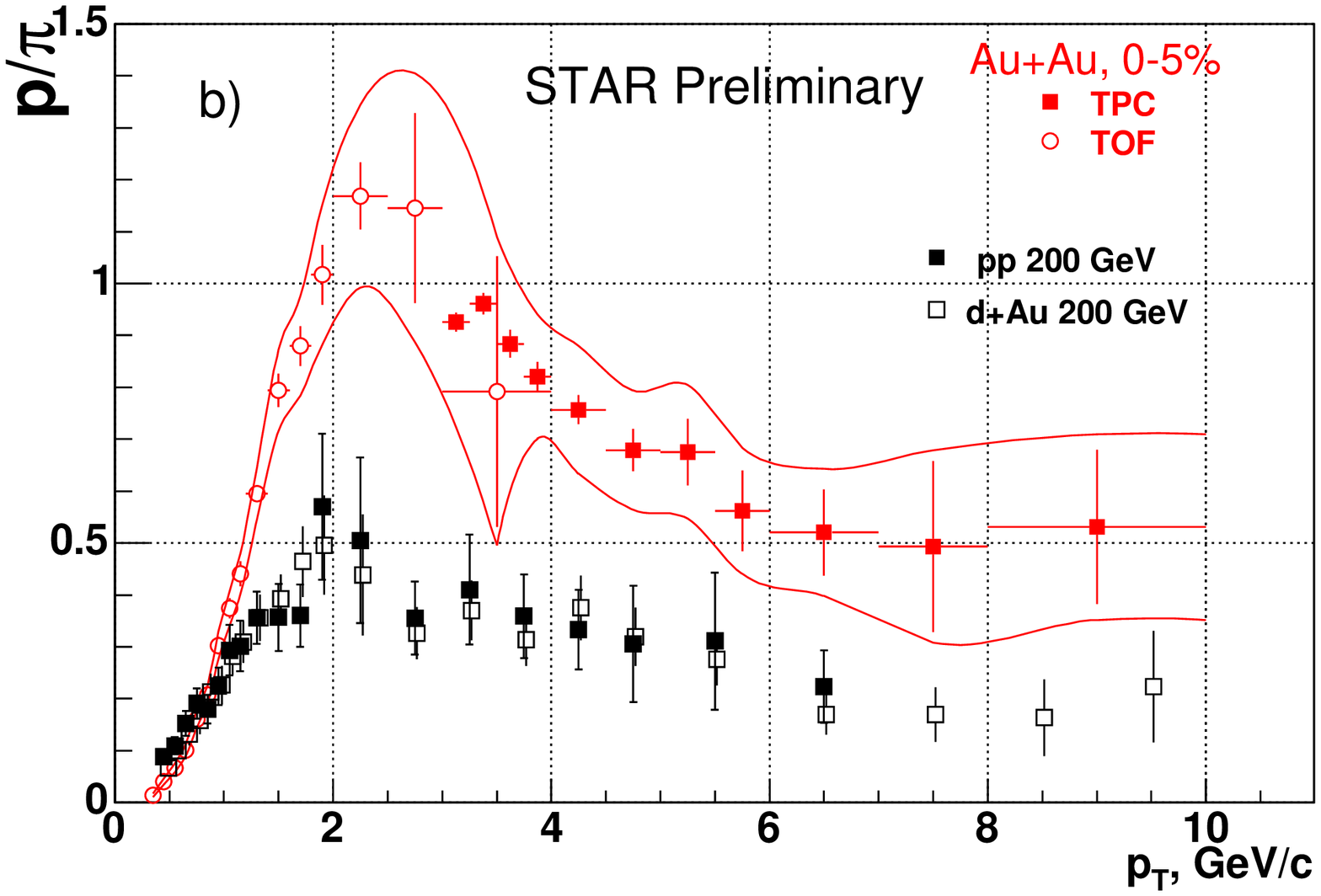,height=2.8in, width=3.5in}
\end{minipage}
\hspace{2.6cm}
\begin{minipage}[b]{60mm}
\epsfig{file=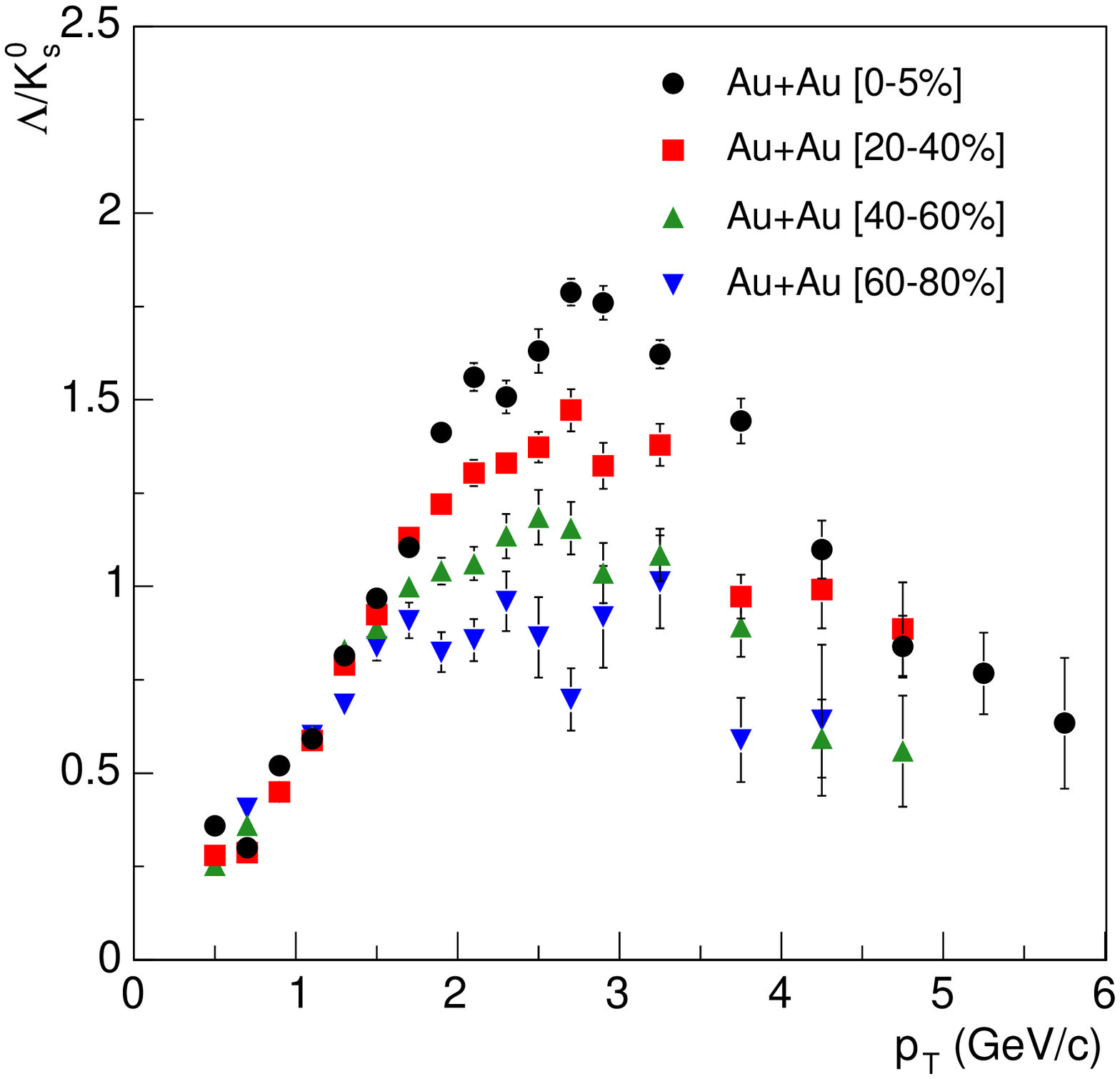,height=3.2in, width=3.2in}
\end{minipage}
\caption{\label{fig:1} Left: Proton-to-pion ratio as a function of transverse momentum
($\pt$) for the 10$\%$ most central Au+Au collisions as well as for p+p and d+Au 
collisions at $\snn = 200$ GeV. The error bars for Au+Au are statistical and the error
bands indicate the systematic uncertainties. For p+p, the error bars are combined 
statistical and systematic.
Right: $\pt$ dependence of the lambda-to-kaon ratio for different centralities in Au+Au 
collisions.}
\end{figure}

Measurements of identified particle yields have shown that baryons are more abundant than 
mesons at intermediate $\pt$ (2--6 GeV/$c$) in heavy-ion collisions. This is illustrated 
in Figure~\ref{fig:1} where both the p/$\pi$ (left) and $\Lambda$/K$^0_{\rm s}$ ratio (right) 
is shown as a function of transverse momentum~\cite{lambdakaon,protonpion}. Both ratios 
increase up to approximately 3~GeV/$c$ and exceed unity for the most central collisions. 
The p/$\pi$ ratio in p+p and d+Au collisions is shown for comparison, with values
below 0.5 for the entire $\pt$ range. 
If fragmentation is the dominant production mechanisms, one would expect the p/$\pi$ ratio 
in Au+Au collisions to be similar to p+p collisions. The observed enhancement in central 
Au+Au collisions therefore indicates that additional non-perturbative effects dominate at 
intermediate $\pt$. 
From the present statistics it seems that vacuum fragmentation is approached in Au+Au collisions 
above 6~GeV/$c$. In addition, the $\Lambda$/K$^0_{\rm s}$ ratio exhibits a continuous evolution 
from peripheral to central collisions.

\section{Particle azimuthal correlations}
Due to full azimuthal coverage the STAR detector has good capabilities to study leading 
particle correlations. Recent STAR results on di-hadron azimuthal correlations at 
high-$\pt$ are reviewed in this section. 

It was observed that jet-like azimuthal angular ($\dphi$) correlations opposite to trigger 
jets in the $\pt$ range $4<\pttrig<6$~GeV/$c$ are suppressed for associate particles 
selected in $2<\ptasso<4$~GeV/$c$ in central Au-Au collisions, whereas no suppression 
was measured in p+p, d+Au and peripheral collisions~\cite{stardAuSpectraCorr}.
In the presence of final state energy loss the observation of high-$\pt$ hadrons is biased
towards production close to the surface. This surface bias is likely different in
correlations where in particular the away side hadron should be more sensitive to medium
properties.
Since momentum has to be conserved the fragmentation products of the away-side jet cannot 
disappear. Instead the $\pt$ spectrum of these particles is softened and the corresponding
$\dphi$ distribution using the low-$\pt$ associate particles is similar to a momentum 
balanced shape~\cite{lowPt}.


\begin{figure}[t]
\begin{minipage}[t]{60mm}
\hspace{-4mm}
\epsfig{file=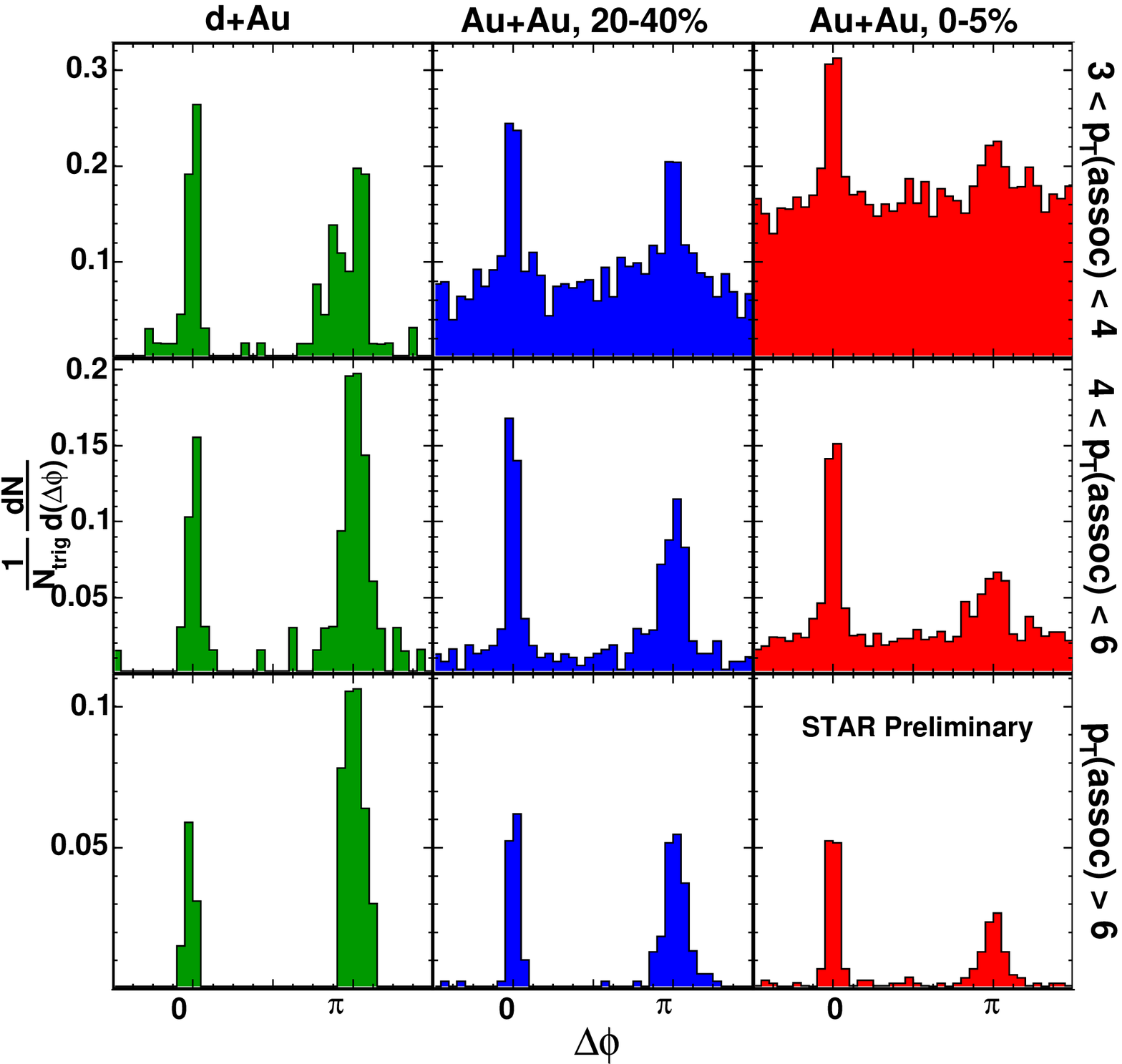,height=3.6in, width=3.5in}
\end{minipage}
\hspace{2.7cm}
\begin{minipage}[b]{60mm}
\epsfig{file=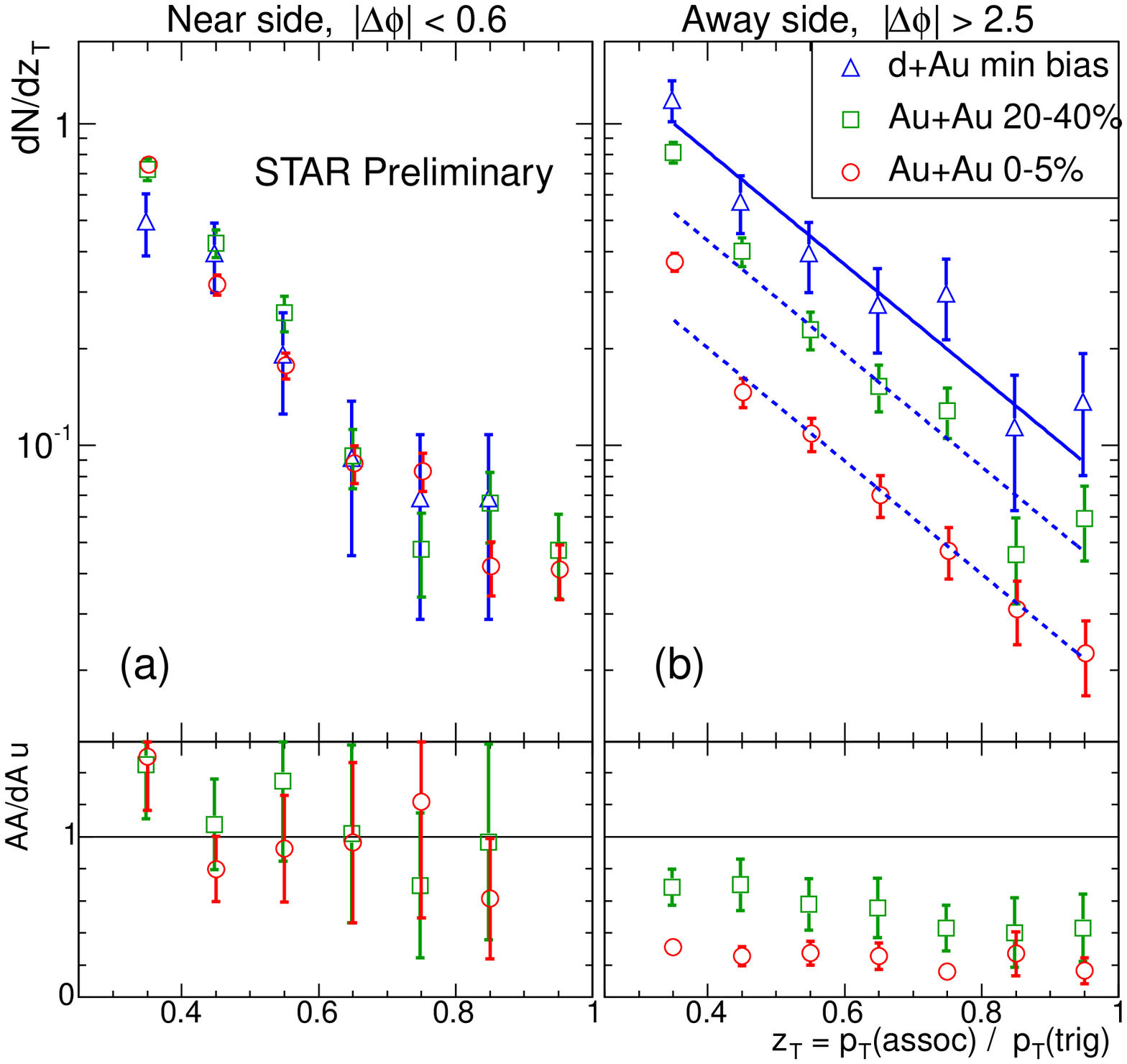,height=3.3in, width=3in}
\end{minipage}
\caption{\label{fig:2} Left: Azimuthal correlation distribution of high-$\pt$ charged
hadrons selected in the transverse momentum range $8 < \pttrig <15$ GeV and associated 
with charged hadrons of three different $\pt$ ranges in d+Au and Au+Au collisions 
(two different centralities) at $\snn = 200$ GeV. The distributions are not corrected
for background contributions.
Right: Trigger-normalized charged hadron fragmentation function ($dN/d\zt$) for near-
(left) and away-side (right) correlations in d+Au and Au+Au collisions. The lines are 
described in the text. The lower panels show the ratio of $dN/d\zt$ for Au+Au relative
to d+Au collisions.}
\end{figure}

The high statistics Au+Au run allows to extend the di-hadron azimuthal correlation analysis
to much higher $\pt$ thresholds for the trigger and associated particle~\cite{dan}. 
Figure~\ref{fig:2}, left, shows the azimuthal correlation distribution of high-$\pt$ charged 
hadrons selected in the transverse momentum range $8 < \pttrig <15$ GeV and associated with 
charged hadrons of three different $\pt$ ranges in d+Au and Au+Au collisions (two different 
centralities) at $\snn = 200$ GeV. No background subtraction is applied to the distributions.
A clear back-to-back peak is observed for all collision systems. However, this peak is
suppressed in Au+Au collisions for all associate $\pt$ ranges.
This is shown in more detail in Figure~\ref{fig:2}, right, where the di-hadron
fragmentation function of charged hadrons is plotted for the near- (left) and away-side 
(right) correlations in d+Au and Au+Au collisions as a function of $\zt~(= \ptasso/\pttrig$),
as proposed in reference~\cite{XNW04}. 
The near-side peak shows no significant modification from d+Au to central Au+Au whereas the 
away-side peak is strongly suppressed in Au+Au collisions. The solid line in Figure~\ref{fig:2}, 
right, is an exponential fit to the d+Au data points. The dashed lines represent the same 
exponential fit scaled down by a factor of 0.54 and 0.25 to approximate the Au+Au yields in 
the centrality bins 20--40$\%$ and 0--5$\%$, respectively. The suppression in central Au+Au 
shows essentially no $\zt$ dependence above approximately 0.3.
The size of the suppression, a factor 4--5, is similar to the suppression factors measured 
for inclusive particle spectra at high-$\pt$ ($\pt\gtrsim$ 6 GeV/$c$). 
Remarkably, despite the suppression, the slope of the fragmentation distribution is the
same in d+Au and central Au+Au collisions. 

In a multiple scattering scenario, one would intuitively expect that the strong suppression  
also lead to a broadening of the away-side peak~\cite{rolf,vitev}. Such a broadening is not 
observed in central Au+Au collisions. Instead the width of the away-side peak is approximately 
independent of centrality. 

\section{Summary}
In this paper recent high-$\pt$ results from the STAR experiment are presented. The large 
enhancement of the particle ratios p/$\pi$ and $\Lambda$/K$^0_{\rm s}$ in Au+Au collisions 
indicates that vacuum fragmentation is not the dominant source of particle production at 
intermediate $\pt$.
Di-hadron azimuthal correlations at higher $\pt$ exhibit a finite suppression of the
back-to-back peak whereas the width of the away-side peak is similar for peripheral and 
central Au+Au collisions. 
These results provide stringent constraints for parton energy loss models and the medium
density~\cite{dan}.

\section*{Acknowledgments}
A. M. thanks the Netherlands Organization for Scientific
Research (NWO) for support.
We thank the RHIC Operations Group and RCF at BNL, and the
NERSC Center at LBNL for their support. This work was supported
in part by the Offices of NP and HEP within the U.S. DOE Office 
of Science; the U.S. NSF; the BMBF of Germany; IN2P3, RA, RPL, and
EMN of France; EPSRC of the United Kingdom; FAPESP of Brazil;
the Russian Ministry of Science and Technology; the Ministry of
Education and the NNSFC of China; IRP and GA of the Czech Republic,
FOM of the Netherlands, DAE, DST, and CSIR of the Government
of India; Swiss NSF; the Polish State Committee for Scientific 
Research; SRDA of Slovakia, and the Korea Sci. $\&$ Eng. Foundation.

\end{document}